# *Gamma band oscillations reflect sensory and affective dimensions of pain*

Yuanyuan Lyu[1, 2, *], Francesca Zidda[1], Stefan Radev[1], Hongcai Liu[1], Xiaoli Guo[2], Shanbao Tong[2], Herta Flor[1], Jamila Andoh[1, *]

[1]Department of Cognitive and Clinical Neuroscience, Central Institute of Mental Health, Medical Faculty Mannheim, Heidelberg University, Mannheim 68159, Germany.

[2]School of Biomedical Engineering, Shanghai Jiao Tong University, Shanghai 200240, China.

**Conflict of Interest:** None.

**Acknowledgements:** This work was supported by a grant from EFIC Grünenthal (EGG) to JA, a grant from the Deutsche Forschungsgemeinschaft (SFB1158/B07) to HF and JA, and a fellowship by the China Scholarship Council (CSC) to YL and HL. We also thank Daniel Wagner for providing a picture for illustration purposes for Figure 1.

*Correspondence should be addressed to Jamila Andoh, email: jamila.andoh@zi-mannheim.de and Yuanyuan Lyu, email: lvyuanyuanoo@gmail.com.

*Abstract*

Pain is a multidimensional process, which can be modulated by emotions, however, the mechanisms underlying this modulation are unknown. We used pictures with different emotional valence (negative, positive, neutral) as primes and applied electrical painful stimuli as targets to healthy participants. We assessed pain intensity and unpleasantness ratings and recorded electroencephalograms (EEG). We found that pain unpleasantness, and not pain intensity ratings were modulated by emotion, with increased ratings for negative and decreased for positive pictures. We also found two consecutive gamma band oscillations (GBOs) related to pain processing from time frequency analyses of the EEG signals. An early GBO had a cortical distribution contralateral to the painful stimulus, and its amplitude was positively correlated with intensity and unpleasantness ratings, but not with prime valence. The late GBO had a centroparietal distribution and its amplitude was larger for negative compared to neutral and positive pictures. The emotional modulation effect (negative versus positive) of the late GBO amplitude was positively correlated with pain unpleasantness. The early GBO might reflect the overall pain perception, possibly involving the thalamocortical circuit, while the late GBO might be related to the affective dimension of pain and top-down related processes.

*Significance statement*

Pain experience can be modulated by emotions. The cortical representation of such modulation is however still under debate. We found that emotional valence modulated pain unpleasantness and not intensity ratings. Using electroencephalography, time frequency analyses showed two consecutive gamma band oscillations (GBOs) related to pain processing: an early GBO, reflecting the overall pain perception; and a late GBO, reflecting emotional modulation in the affective dimension of pain. These GBO are in line with a serial model of pain perception and might reflect a relevant neural marker for chronic pain conditions.

**Introduction**

Pain is an unpleasant sensory and emotional experience associated with potential or actual tissue damage or described in such term. From this definition, it emerges that pain contains both a sensory-discriminative and an affective-motivational dimension (Price 2002; Bushnell et al. 2013). The sensory-discriminative dimension refers to the intensity quality of pain, whereas the affective-motivational dimension reflects the unpleasantness of a painful experience and the associated tendency to avoid it (Melzack and Casey 1968; Kerns et al. 1985; Arnold et al. 2008). Although pain intensity and unpleasantness ratings are known to be highly correlated, experimental manipulations using various modalities (visual, auditory, olfactory) showed a differential modulation of the two dimensions. For instance, pleasant compared with unpleasant odors, could decrease pain unpleasantness but had little effect on pain intensity (Villemure et al. 2003; Villemure and Bushnell 2009). Listening to pleasant music, however, reduced both pain intensity and unpleasantness (Roy et al. 2008). In all these studies, presentations of emotional material and painful stimulation occurred simultaneously. Additionally, these studies used a relatively long trial duration (> 6s), which might introduce cognitive confounds to the emotional modulation of pain, such as attentional processes. Thus, special experimental paradigm, like prime-target presentation, might be useful to reduce those attentional or cognitive factors on emotional modulation of pain.

Cortical oscillations, which can be extracted by frequency domain analysis from scalp EEG signal, reflect synchronization of neuronal ensembles (Pfurtscheller and Da Silva 1999). Recently, a focus was put on the cortical oscillations related to pain (Ploner et al. 2017), such as the lower bands, like alpha (8 – 13 Hz), beta (14 – 30 Hz) and also higher gamma band oscillations (30 – 100 Hz, GBOs). For instance, the amplitude of GBO has been shown to be closely coupled with the perceived pain intensity, rather than the actual stimulus intensity (Gross et al. 2007; Zhang et al. 2012; Schulz et al. 2015; Nickel et al. 2017), suggesting that GBO could reflect the sensory-discriminative dimension of pain. However, it remains controversial whether GBO also carry information about the affective dimension of pain perception and thus changes in emotional valence could also affect GBO (Senkowski et al. 2011; Hauck et al. 2013; Hauck et al. 2015; Tiemann et al. 2015; Nicolardi and Valentini 2016).

In the present study, we investigated the influence of emotional valence on pain perception using both subjective pain ratings and cortical oscillatory measures. We presented pictures of various types of emotional valence (negative, neutral, positive) as primes and then applied painful electrical stimuli to healthy participants. Changes in pain perception were assessed using pain intensity and unpleasantness ratings. We hypothesized that emotional valence would modulate pain ratings, for example, negative pictures would increase and positive pictures should decrease pain perception compared to neutral pictures. For cortical oscillations, we expected that the amplitude of GBO would be positively correlated with pain ratings, and would be also modulated by emotional valence, especially for the negative one because of its stronger adaptive value.

## Materials and Methods

### *Participants*

Twenty-one healthy subjects (age: 23.5 ± 2.6, 11 females) participated in the present study. Participants were all right-handed (mean score of the sample = +95.6), as assessed using the Edinburgh Handedness Inventory (Oldfield 1971), and had no history of mental or neurological disorders. The participants were informed about the purpose and the methods used in the study and signed informed consent. The study was approved by the Ethics Committee of the Medical Faculty Mannheim of Heidelberg University.

*Experimental procedure*

The participants sat in a comfortable chair in front of a monitor and the distance between the eyes and the monitor was approximately 50 cm. Before each trial, a fixation cross was presented in the center of a gray background for a randomized duration between 1200 ms and 2400 ms denoting the inter-trial interval (Figure 1). Following the first fixation cross, a prime picture was displayed for 200 ms and was then replaced by a second fixation cross. After 200 ms a painful electrical stimulus was applied at the left forearm by a bar electrode. Then, after 1000 ms from the onset of electrical stimulation, the participants were asked to perform ratings on two consecutive visual analog scales (VAS): the first VAS related to the intensity of pain (i.e., How intense was the painful stimulus?) ranging from no pain to most intense pain imaginable; the second VAS was used to rate the unpleasantness of pain (i.e., How unpleasant was the stimulus?) and ranged from not at all unpleasant to most unpleasant pain imaginable. The prime pictures contained emotions of different valence (negative, neutral, or positive), and were taken from the International Affective Picture System (Lang et al. 2008) [1]. The pictures were selected based on normative ratings on the dimensions of affective valence (negative: 2.17 ± 0.36, neutral: 5.22 ± 0.55, positive: 7.40 ± 0.40) and arousal (negative: 5.74 ± 0.51, neutral: 4.27 ± 0.59, positive: 4.83 ±

---

[1] Picture numbers were: neutral (1390, 1903, 2025, 2032, 2235, 2372, 2487, 2514, 2521, 5900, 6000, 7011, 7013, 7018, 7021, 7033, 7042, 7044, 7057, 7058, 7077, 7081, 7096, 7137, 7140, 7183, 7184, 7188, 7237, 7248, 7249, 7513, 7550, 7560, 7620, 7632, 7820, 7830, 9150, 9468), positive (1410, 2035, 2045, 2050, 2057, 2070, 2150, 2274, 2311, 2340, 2352, 1440, 2360, 2395, 2550, 2660, 4640, 4641, 5220, 5480, 5825, 5830, 1463, 7230, 7260, 7270, 7330, 7470, 8120, 8461, 8496, 8501, 8502, 1510, 8540, 1630, 1710, 1721, 1750, 1999), negative (2301, 2352, 2710, 2800, 2900, 2981, 3016, 3017, 3051, 3059, 3061, 3064, 3168, 3181, 3185, 3220, 3225, 3301, 3550, 6022, 6213, 6243, 6415, 6520, 6560, 6563, 9040, 9043, 9075, 9140, 9181, 9185, 9253, 9265, 9332, 9405, 9571, 9635, 9800, 9902).

0.73), and the rating scale ranged from 1 to 9, with 1 representing low pleasure and low arousal and 9 representing high pleasure and high arousal (Lang et al. 1997). We converted the rating scales to 0 - 100 for analysis. Although the arousal ratings of valence were different, we analyzed the results for a subset of stimuli with comparable arousal to show that arousal is not the main contributor for the present results, see the discussion. The three valence conditions consisted of 40 pictures each and each picture was only presented once, i.e. 120 trials in total (40 × 3). The electrical stimuli were generated by a constant stimulator (Digitimer® DS7A, United Kingdom). For each participant, we measured the perceived perception threshold, pain threshold and pain tolerance three times before the experiment, respectively. To make the electrical stimulus painful but tolerable, the chosen stimulation intensity was defined as mean pain threshold plus 80% of the difference between the pain tolerance and mean pain threshold. We applied rectangular pulses at random durations between 3 to 7 ms to increase the variability of the pain ratings.

----------------------

Insert Figure 1 around here

----------------------

*EEG Acquisition and Analysis*

The EEG signals were amplified by BrainAmp amplifiers (BrainProducts GmbH, Munich, Germany) and collected with BrainVision Recorder software, sampled at 1000 Hz and filtered online between 0.016 Hz and 250 Hz. EEG was recorded using a 64-channel actiCap with active Ag/Agcl electrodes. Electrode positions on the cap were following the standard 10-10 system. Two more electrodes were used to record vertical and horizontal electro-oculograms to detect eye movements and blinks. The ground electrode was placed at AFz and the reference electrode was placed at FCz. Electrode impedance was kept at less than 20 kΩ as suggested from the manufacturer. The active electrodes used here were demonstrated to be insensitive to moderate levels of impedance (< 50 kΩ) when compared to passive electrodes for measurements like EEG spectra (Mathewson et al. 2017).

EEG data were preprocessed using EEGLAB 15.3.6 (Delorme and Makeig 2004). Data were first filtered using a 1Hz high-pass filter and then interpolated the bad channels (percentage: 2.71 ± 2.06%). The filtered data were re-referenced to an average reference except for the eye electrodes and segmented in epochs from 1 second before to 2 seconds after the onset of the prime picture. Epochs with motion artefacts were rejected by visual inspection and the behavioral data of the rejected epochs were also excluded. Independent component analysis was applied to the clean epoched data and components representing artefactual non-brain activity were rejected, i.e. eye movements, cardiac activity, powerline noise (50Hz) and electrical stimulation artefacts. Then the preprocessed epochs were assigned to the three conditions based on the picture valence (negative, neutral, positive).

Event-related spectral perturbation (ERSP) analyses (Makeig 1993) were performed using the *newtimef()* function in EEGLAB. Morlet wavelets transformation was applied to each single EEG epoch with a sliding window. The window had a length of 1115 points (1115 ms) and was shifted in a step of 1 data point (1 ms). The frequency range was from 3 Hz to 100 Hz with a resolution of 1Hz. The cycles of wavelets increased linearly from 3 cycles at the lowest frequency (3 Hz) to 20 cycles at the highest (100 Hz) to achieve a good trade-off between the time and frequency resolutions (McLelland et al. 2016).

The time-frequency transformed data were averaged across trials for each condition and each subject. The ERSP amplitude was calculated as $10*\log_{10}$ transformed multiples of amplitude change with respect to the baseline. The baseline was defined as the 442 time points before the prime pictures. Global grand averaged ERSPs were obtained by averaging ERSPs across all prime pictures and all participants. After visual inspection, we found two prominent GBOs with increased amplitude after painful electrical stimulus in the stimulation in the following time-frequency windows and regions, i.e., 1) early GBO, 420 – 500 ms, 35 – 70 Hz, right centroparietal area (FCz, FC2, FC4, Cz, C2, C4, CPz, CP2, CP4, Pz, P2, P4); 2) late GBO, 500 – 660 ms, 60 – 95 Hz, middle centroparietal area (C3, C1, Cz, C2, C4, CP3, CP1, CPz, CP2, CP4, P3, P1, Pz, P2, P4). For further analysis, the amplitude of the each GBO was calculated by averaging the ERSP amplitudes across the above window and region for each participant and each prime valence. The GBO contains both phase-locked and non-phase-locked components, denoting as total GBO hereafter. Meanwhile, the inter-trial coherence (ITC) (Delorme and Makeig 2004), also known as event-related phase-locking value, was calculated for each GBO.

To investigate the non-phase-locked component of GBOs, we remove the ERP signal from the EEG segments and calculate the induced ERSP use the same parameters as we calculated total ERSP. Then we extracted the early and late induced GBO from the same time-frequency-channel window for later statistical analysis.

*Statistical Analysis*

The pain ratings (intensity and unpleasantness) and ERSP values in different time-frequency windows were analyzed using one-way repeated measures analyses of variance (ANOVA) with prime valence (negative, neutral and positive) as a within-subject factor. To test whether the early and late

GBOs shared the same characteristics of phase locking activity, ITC values were analyzed using a 2 × 3 repeated measures ANOVA, taking prime valence (negative, neutral and positive) and GBO (early and late) as within-subject factors. Post hoc tests were corrected for multiple comparisons using Bonferroni corrections. We also examined the relationship between the pain intensity and unpleasantness ratings and ERSPs using Spearman's rank correlations. Although previous studies showed linear trends for increased pain ratings with negative valence of the presented pictures (Rhudy et al. 2006; Kenntner-Mabiala et al. 2008), they also observed that only the neural activity for the negative valence differed significantly from the positive or the neutral in the N150 component of ERP (event related potential) and spinal nociceptive response after pain (Kenntner-Mabiala and Pauli 2005; Roy et al. 2009; Roy et al. 2011). Thus, to quantify this negative emotional modulation effect (i.e. negative versus neutral and also negative versus positive), we performed transformations on pain ratings and GBO amplitude and correlation analyses between them. Pain ratings (i.e., intensity, unpleasantness) were normalized by dividing them between negative and neutral prime valence ($INT_{(neg/neu)}$, $UNP_{(neg/neu)}$) and between negative and positive prime valence ($INT_{(neg/pos)}$, $UNP_{(neg/pos)}$). For the GBO, since their amplitude was log-transformed, normalization was performed by subtracting GBO amplitude between the neutral and the negative prime valence ($GBO_{(neg-neu)}$), and between the positive and the negative prime valence ($GBO_{(neg-pos)}$). Outliers were detected using the interquartile range (IQR), defined as the upper quartile minus the lower quartile. Values outside the range of the lower quartile - 1.5*IQR to the upper quartile + 1.5*IQR were excluded from all analyses. The significance level was set at $p < 0.05$. All data are presented as means ± standard deviation.

**Results**

*Pain intensity and unpleasantness ratings*

Pain intensity ratings were comparable across valence conditions ($F(2, 40) = 1.10$, $p = 0.34$, negative: 31.49 ± 18.31, neutral: 31.70 ± 16.28, positive: 30.41 ± 17.39).

In contrast, there was a main effect of prime valence on pain unpleasantness ratings ($F(2, 40)= 15.85$, $p < 0.001$). Post hoc tests indicated that pain unpleasantness ratings were significantly higher for the negative (36.62 ± 19.11) than the neutral (32.15 ± 18.24, $p= 0.001$) and the positive (29.68 ± 18.52, $p= 0.002$) prime valence. In addition, pain unpleasantness ratings were significantly higher for the neutral than for the positive ($p= 0.023$) prime valence, see Figure 2A.

We also found a positive correlation between the pain intensity and unpleasantness ratings (rho= 0.851, $p< 0.001$, N= 19, outliers: participants 3, 8), Figure 2B.

----------------------

Insert Figure 2 around here

----------------------

*Total gamma band oscillations*

After visual inspection, we found two prominent GBOs following the painful electrical stimuli (Figure 3A). An early GBO (35 - 70 Hz) appeared in 20 – 100 ms after the electrical stimulus, centrally distributed in the hemisphere contralateral to the location of the stimulus application (Figure 3B). A late

GBO, in a higher gamma band (60 – 95 Hz), appeared in 100 – 260 ms after the electrical stimuli, with a centroparietal distribution (Figure 3C).

----------------------

Insert Figure 3 around here

----------------------

The amplitude of the early GBO was comparable across prime valences ($F(2, 40)= 1.348$, $p = 0.271$, negative: $0.60 \pm 0.51$ dB, neutral: $0.45 \pm 0.57$ dB, positive: $0.53 \pm 0.59$ dB), see Figure 3D. In addition, the mean amplitude of the early GBO across valence conditions was positively correlated with the mean pain intensity rating across valence conditions (Figure 4A, rho= 0.608, p= 0.009, N= 18, outliers: participants 3, 6, 8) and with the mean pain unpleasantness rating across valence conditions (Figure 4B, rho= 0.558, p= 0.015, N= 19, outliers: participants 3, 6). Since there was no emotional modulation effect showed in the amplitude of early GBO, we did not perform correlation analyses between normalized amplitude of GBO and pain ratings.

----------------------

Insert Figure 4 around here

----------------------

The amplitude of the late GBO revealed a main effect of prime valence ($F(2, 40)= 5.877$, $p= 0.006$, Figure 3E). Post hoc tests indicated that the amplitude of the late GBO for the negative prime valence ($0.66 \pm 0.52$ dB) was larger than that for the neutral ($0.45 \pm 0.44$ dB, p= 0.027) and positive ($0.50 \pm 0.48$ dB, p= 0.046) primes. In addition, the amplitude of the late GBO was comparable between neutral and positive prime valence (p= 1.00). However, unlike the early GBO, the mean amplitude of the late GBO across valence conditions was not significantly correlated with the mean pain intensity ratings (rho= -0.018, p = 0.943, N= 19, outliers: participants 3, 8) nor with the mean pain unpleasantness ratings (rho= 0.060, p= 0.797, N= 21, no outliers).

Correlation analyses showed that there was no relationship between the normalized late $GBO_{(neg-neu)}$ amplitude and $UNP_{(neg/neu)}$ (rho= 0.270, p= 0.262, N= 19, outliers: participants 5, 14). However, the normalized late $GBO_{(neg-pos)}$ amplitude was significantly positively correlated with $UNP_{(neg/pos)}$ (rho= 0.511, p= 0.027, N= 19, outliers: participants 5, 12), see Figure 5.

----------------------

Insert Figure 5 around here

----------------------

Finally, ITC values exhibited a significant main effect of GBO (F(1, 20)= 27.520, p <0.001) but no significant main effect of prime valence (F(2, 40)= 0.384, p= 0.683), and no the interaction between GBO and prime valence (F(2, 40)= 1.544, p= 0.226). The early GBO (0.23 ± 0.06) was more phase locked than the late GBO (0.16 ± 0.02), Figure 6.

----------------------

Insert Figure 6 around here

----------------------

*Subset analysis on positive and negative primes with comparable arousal ratings*

The normative ratings of arousal differed between the negative and positive pictures we selected for this study (F(2, 117) = 58.04, p < 0.001). To assess the potential confound of arousal, we chose a subset of 26 positive and 26 negative pictures from our original dataset (40 positive and 40 negative pictures), with comparable arousal values (T(46.8) = 1.77, p = 0.08). Similar to our main findings, we observed that compared with positive pictures, the negative pictures increased pain unpleasantness (T(20) = 3.37, p = 0.003) and not pain intensity ratings (T(20) = 0.52, p = 0.612). Regarding GBOs, we also found a significant main effect of prime valence on the late GBO (T(20) = 3.15, p = 0.005), indicating that the amplitude for the late GBO for the negative valence were larger than the positive ones. No effect of prime valence was found for the early GBO (T(20) = 0.94, p= 0.358), which correspond well to the above results.

*Induced gamma band oscillations*

For the early GBO amplitude, ANOVA showed an insignificant main effect of prime valence (F(2, 40)= 1.374, p = 0.265). Meanwhile, the early GBO amplitude was not significantly correlated with neither pain intensity rating (rho = 0.41, p = 0.63) nor pain unpleasantness rating (rho = 0.47, p = 0.19). Thus, the induced early GBO cannot reflect the encoding of pain perception.

For the late GBO amplitude, ANOVA showed a significant main effect of prime valence (F(2, 40)= 6.547, p = 0.003). Post-hoc analysis showed the late GBO amplitude after negative prime (0.66 ± 0.52 dB) was significantly larger than the one after positive prime (0.44 ± 0.45 dB, p= 0.035) and neutral prime (0.50 ± 0.47 dB, p= 0.019). The normalized late $GBO_{(neg-pos)}$ amplitude was significantly positively correlated with UNP$_{(neg/pos)}$ (rho= 0.519, p= 0.024, N= 19, outliers: participants 5, 12). Thus, the induced late GBO still reflect the emotional modulation.

**Discussion**

We investigated how the sensory and affective dimensions of pain were modulated by emotional valence using self-reports of pain and gamma band neural oscillations. Pain ratings showed that emotional valence affected pain unpleasantness but not pain intensity. Negative prime pictures increased pain unpleasantness, while positive prime pictures decreased it. Moreover, we identified two consecutive GBOs following painful stimuli. The early GBO correlated with the overall pain intensity and pain unpleasantness ratings and was not influenced by emotional valence. On the other hand, the late GBO in the higher gamma band was modulated by emotional valence, particularly for the negative valence condition.

Only the pain unpleasantness ratings were significantly different across the three prime valences, indicating that the affective rather than the sensory dimension of pain was sensitive to the emotional pictures. The visual stimuli used in the current design were characterized by two dimensions: valence and arousal, but the modulation effect is most likely driven by the dimension of valence. First, in the subset analysis on positive and negative pictures with comparable arousal ratings, the negative pictures elicited significant larger unpleasantness ratings than positive ones. Second, according to the distraction theory, the pictures with high arousal rating (positive/negative) would trigger a decrease in pain perception than neural ones, which is not the case in our results (Buffington et al. 2005; Wiech et al. 2005; Dunckley et al. 2007). Our results add a new perspective to the current literature and the experimental design used in this study was intended to optimize the assessment of emotional modulation of pain. On one hand, the painful stimuli used in the present study were delivered without additional concomitant confounds. This may limit the interaction of additional cognitive factors such as attention as concomitant presentation of emotional stimuli with delivery of painful stimuli was often used in previous studies (Kenntner-Mabiala et al. 2008; Villemure and Bushnell 2009; Senkowski et al. 2011; Ring et al. 2013). On the other hand, the assessment of both intensity and unpleasantness pain ratings enabled us to differentiate between sensory and affective dimensions of pain. Indeed, some studies have not assessed both pain unpleasantness and intensity ratings (Godinho et al. 2006; Roy et al. 2011; Senkowski et al. 2011), and might have merged the sensory and affective dimensions of pain. Our findings are in line with the neuroimaging literature highlighting the need to assess both sensory and affective dimensions of pain (Price 2002; Tracey and Mantyh 2007; Bushnell et al. 2013).

Regarding GBOs, we were able to show two consecutive GBOs following painful stimuli and several dissociations between them: the early GBO had a distribution widespread over the contralateral S1, while the late GBO were widespread over a large centroparietal area in the midline and appeared in a higher gamma band and at later time window. Regarding the behavioral measures, the early GBO encoded the overall perceived pain intensity and unpleasantness while the late one was modulated by emotional valence. Moreover, the phase-locked value of the early GBO was significantly larger than the late one. When we only considering the induced GBO, the correlations between early GBO and overall pain ratings did not reach significant level, while modulatory effect and the relationship with unpleasantness rating still hold in the late GBO. Thus, the early and late GBO mainly originate from phase-locked component and non-phase-locked component, respectively, and might be mediated by different mechanisms.

The early GBO is mostly likely a time-frequency representation of the early complex N20-P30 wave of somatosensory evoked potential (SEP) elicited by the electrical stimulation of the upper limb. The N20-P30 is phase-locked and originate from the contralateral somatosensory cortex (Cruccu et al. 2008). Previous study showed the median-nerve SEP contained oscillation components ranging from 30 – 80 Hz (Chen and Herrmann 2001). Our result showed the phase-locked component of early GBO encode the perceived pain intensity and unpleasantness in the phasic experimental pain condition. Such findings are not surprising since pain intensity and unpleasantness ratings were highly correlated. The early GBO may reflect the temporal binding of thalamocortical projections (Price 2002; Bruno and Sakmann 2006). Simultaneous recordings from the ventral posterior medial nucleus of the thalamus and corresponding cortical columns showed that the thalamic GBO had a strong phase modulation to the cortical GBO evoked by brief single-whisker deflection in rats (Minlebaev et al. 2011). Likewise, source analysis of magnetoencephalographic data in humans showed such coherent thalamocortical GBO in the auditory modality (Ribary et al. 1991). Furthermore, our results showed that the early GBO was not significantly modulated by emotional valence.

Unlike previous reports indicating that GBO encode the perceived pain intensity in a phasic pain condition (Gross et al. 2007; Zhang et al. 2012; Hauck et al. 2013; Schulz et al. 2015), our results showed that the late non-phase-locked GBO did not directly encoded the pain perception but was modulated by emotion valence. The direct comparison of amplitude of the late GBO among the different prime valence revealed that increased response to negative than positive and neutral prime valence. The role of stimuli valence, especially negative items has previously been shown to affect GBO in a passive viewing mode (Headley and Paré 2013). Our results indicated that the negative valence from priming visual stimuli could also induce a higher GBO later in pain perception process and may reflect a top-down modulation.

Likewise, a EEG study presenting pain stimuli together with emotional facial expressions also showed an emotional modulation of GBO, in which the authors found facial expression fear elicited increased GBO compared with facial expression angry (Senkowski et al. 2011). Since synchrony in gamma band is related to the communication between cortical areas (Fries 2005), it can be speculated that the increased late GBO in the centroparietal area may represent upregulated descending pain processing pathway triggered by negative prime. Such a top-down modulation may also contribute to the increased pain unpleasantness rating. Moreover, the emotional modulation effect from negative to positive of the late GBO is significantly correlated to that of pain unpleasantness ratings. Negative affects facilitate avoidance-motivated behavior while positive affects facilitate approach-motivated behavior (Cacioppo and Berntson 1994). As an aversive stimulus, acute pain also triggers avoidance-motivated behavior (Bromm and Lorenz 1998). The late GBO might represent the avoidance-motivated behavior, as negative prime and pain would enhance the effect while positive prime and pain would counteract it. Overall, the late GBO might reveal the emotional modulation in the affective dimension of pain perception.

Finally, our results are in agreement with a serial model of pain perception (Price 2000), as the early GBO seems to encode the overall pain intensity and unpleasantness but the late GBO indicates the emotional modulation in the affective dimension occurs later. Early GBO would be fundamental to the late GBO. Further studies are needed to clarify the mechanisms underlying the GBOs in emotional modulation of pain.

**Limitations**

Some limitations need to be addressed in this study. The duration of the presentation of the pictures was relatively short (200ms) compared with previous studies (2s for (Godinho et al. 2006), 6s for (Kenntner-Mabiala and Pauli 2005; Kenntner-Mabiala et al. 2008)), because we intended to reduce cognitive processing such as attention during picture presentation. Our prime picture duration should, however, have been sufficient since modulatory effects by emotions have been shown to last up to 700ms in an event related potential study (Aguado et al. 2013), which is longer than our prime-target interval (400 ms).

The electrical stimulation used in the current study would inevitably activate the non-nociceptive system while we targeted the nociceptive system. Our results showed the amplitudes of early and late GBOs were associated with pain ratings, indicating the brain response following electrical stimulation carries nociceptive information. In future studies, the use of laser stimulation or intraepidermal electrical stimulation might be preferable, this would selectively or largely preferentially activate cutaneous Aδ- and C-fiber nociceptors (Inui et al. 2002; Plaghki and Mouraux 2003). Alternatively, the use of nonpainful electrical stimulation as a control condition might also be beneficial.

The GBO following electrical stimulation could be contaminated by the preceding visual-evoked brain activity. To decrease this potential effect, one could also use visual pictures without electrical stimulation as a control condition (Senkowski et al. 2011).

Finally, our results showed that the overall early GBO amplitude was significantly correlated with the overall pain intensity and unpleasantness ratings across the emotional valences, thus, in order to dissociate pain intensity and unpleasantness ratings, we carried out partial correlations. The early GBO was not significantly correlated with the pain intensity rating (p= 0.58) when the unpleasantness rating was controlled for, or the unpleasantness rating (p=0.29) when the intensity rating was controlled for, showing therefore that the two pain dimensions strongly interact with each other and that both might contribute to the early GBO.

**Conclusion**

We showed that emotional valence modulated selectively the affective dimension of pain. Moreover, we observed that an early GBO might reflect the overall sensory discriminative and affective dimensions of pain while the late GBO might reflect the emotional modulation in the affective dimension of pain. Pain perception seems to be composed by serial processes, defined by different temporal dynamics and spatial coding.

*Abbreviations:* Visual analog scale (VAS), Electroencephalography (EEG), Event-related spectral perturbation (ERSP), Interquartile range (IQR), Gamma band oscillation (GBO).

*References*

Aguado L, Dieguez‒Risco T, Mendez‒Bertolo C, Pozo MA, Hinojosa JA. 2013. Priming effects on the N400 in the affective priming paradigm with facial expressions of emotion. Cogn Affect Behav Neurosci 13:284‒296.

Arnold BS, Alpers GW, Süß H, Friedel E, Kosmützky G, Geier A, Pauli P. 2008. Affective pain modulation in fibromyalgia, somatoform pain disorder, back pain, and healthy controls. European Journal of Pain 12:329‒338.

Bromm B, Lorenz J. 1998. Neurophysiological evaluation of pain. Electroencephalography and clinical neurophysiology 107: 227–253.

Bruno RM, Sakmann B. 2006. Cortex is driven by weak but synchronously active thalamocortical synapses. Science 312: 1622–1627.

Buffington AL, Hanlon CA, McKeown M. 2005. Acute and persistent pain modulation of attention-related anterior cingulate fMRI activations. Pain 113: 172–184.

Bushnell MC, Ceko M, Low LA. 2013. Cognitive and emotional control of pain and its disruption in chronic pain. Nat Rev Neurosci 14: 502–511.

Cacioppo JT, Berntson GG. 1994. Relationship between attitudes and evaluative space: A critical review, with emphasis on the separability of positive and negative substrates. Psychological bulletin 115: 401.

Chen ACN, Herrmann CS. 2001. Perception of pain coincides with the spatial expansion of electroencephalographic dynamics in human subjects. Neuroscience letters 297: 183–186.

Cruccu G, Aminoff M, Curio G, Guerit J, Kakigi R, Mauguiere F, Rossini P, Treede R-D, Garcia-Larrea L. 2008. Recommendations for the clinical use of somatosensory-evoked potentials. Clinical neurophysiology 119: 1705–1719.

Delorme A, Makeig S. 2004. EEGLAB: an open source toolbox for analysis of single-trial EEG dynamics including independent component analysis. J Neurosci Methods 134: 9–21.

Dunckley P, Aziz Q, Wise RG, Brooks J, Tracey I, Chang L. 2007. Attentional modulation of visceral and somatic pain. Neurogastroenterology & Motility 19: 569–577.

Fries P. 2005. A mechanism for cognitive dynamics: neuronal communication through neuronal coherence. Trends in cognitive sciences 9: 474–480.

Godinho F, Magnin M, Frot M, Perchet C, Garcia-Larrea L. 2006. Emotional modulation of pain: is it the sensation or what we recall? Journal of Neuroscience 26:11454–11461.

Gross J, Schnitzler A, Timmermann L, Ploner M. 2007. Gamma oscillations in human primary somatosensory cortex reflect pain perception. PLoS Biol 5:e133.

Hauck M, Domnick C, Lorenz J, Gerloff C, Engel AK. 2015. Top-down and bottom-up modulation of pain-induced oscillations. Front Hum Neurosci 9:375.

Hauck M, Metzner S, Rohlffs F, Lorenz J, Engel AK. 2013. The influence of music and music therapy on pain-induced neuronal oscillations measured by magnetencephalography. Pain 154:539–547.

Headley DB, Paré D. 2013. In sync: Gamma oscillations and emotional memory. Frontiers in behavioral neuroscience 7:170.

Inui K, Tran TD, Hoshiyama M, Kakigi R. 2002. Preferential stimulation of Aδ fibers by intra-epidermal needle electrode in humans. Pain 96:247–252.

Kenntner-Mabiala R, Andreatta M, Wieser MJ, Muhlberger A, Pauli P. 2008. Distinct effects of attention and affect on pain perception and somatosensory evoked potentials. Biol Psychol 78:114–122.

Kenntner-Mabiala R, Pauli P. 2005. Affective modulation of brain potentials to painful and nonpainful stimuli. Psychophysiology 42:559–567.

Kerns RD, Turk DC, Rudy TE. 1985. The west haven-yale multidimensional pain inventory (WHYMPI). Pain 23:345–356.

Lang PJ, Bradley MM, Cuthbert BN. 1997. International affective picture system (IAPS): Technical manual and affective ratings. NIMH Center for the Study of Emotion and Attention:39–58.

Lang PJ, Bradley MM, Cuthbert BN. 2008. International affective picture system (IAPS): Affective ratings of pictures and instruction manual. Technical Report A-8. University of Florida, Gainesville,

FL.

Makeig S. 1993. Auditory event-related dynamics of the EEG spectrum and effects of exposure to tones. Electroencephalography and clinical neurophysiology 86:283–293.

Mathewson KE, Harrison TJ, Kizuk SA. 2017. High and dry? Comparing active dry EEG electrodes to active and passive wet electrodes. Psychophysiology 54:74–82.

McLelland D, Lavergne L, VanRullen R. 2016. The phase of ongoing EEG oscillations predicts the amplitude of peri-saccadic mislocalization. Sci Rep 6:29335.

Melzack R, Casey KL. 1968. Sensory, motivational and central control determinants of pain: a new conceptual model. The skin senses 1.

Minlebaev M, Colonnese M, Tsintsadze T, Sirota A, Khazipov R. 2011. Early gamma oscillations synchronize developing thalamus and cortex. Science 334:226–229.

Nickel MM, May ES, Tiemann L, Schmidt P, Postorino M, Ta Dinh S, Gross J, Ploner M. 2017. Brain oscillations differentially encode noxious stimulus intensity and pain intensity. Neuroimage 148:141–147.

Nicolardi V, Valentini E. 2016. Commentary: Top-down and bottom-up modulation of pain-induced oscillations. Front Hum Neurosci 10:152.

Oldfield RC. 1971. The assessment and analysis of handedness: the Edinburgh inventory. Neuropsychologia 9:97–113.

Pfurtscheller G, Da Silva FL. 1999. Event-related EEG/MEG synchronization and desynchronization: basic principles. Clinical neurophysiology 110:1842–1857.

Plaghki L, Mouraux A. 2003. How do we selectively activate skin nociceptors with a high power infrared laser? Physiology and biophysics of laser stimulation. Neurophysiologie Clinique/Clinical

Neurophysiology 33:269-277.

Ploner M, Sorg C, Gross J. 2017. Brain Rhythms of Pain. Trends Cogn Sci 21:100-110.

Price DD. 2000. Psychological and neural mechanisms of the affective dimension of pain. Science 288:1769-1772.

Price DD. 2002. Central neural mechanisms that interrelate sensory and affective dimensions of pain. Molecular interventions 2:392.

Rhudy JL, Williams AE, McCabe KM, Rambo PL, Russell JL. 2006. Emotional modulation of spinal nociception and pain: the impact of predictable noxious stimulation. Pain 126:221-233.

Ribary U, Ioannides AA, Singh KD, Hasson R, Bolton JP, Lado F, Mogilner A, Llinas R. 1991. Magnetic field tomography of coherent thalamocortical 40-Hz oscillations in humans. Proc Natl Acad Sci U S A 88:11037-11041.

Ring C, Kavussanu M, Willoughby AR. 2013. Emotional modulation of pain-related evoked potentials. Biol Psychol 93:373-376.

Roy M, Lebuis A, Peretz I, Rainville P. 2011. The modulation of pain by attention and emotion: a dissociation of perceptual and spinal nociceptive processes. Eur J Pain 15:641 e641-610.

Roy M, Peretz I, Rainville P. 2008. Emotional valence contributes to music-induced analgesia. Pain 134:140-147.

Roy M, Piche M, Chen JI, Peretz I, Rainville P. 2009. Cerebral and spinal modulation of pain by emotions. Proc Natl Acad Sci U S A 106:20900-20905.

Schulz E, May ES, Postorino M, Tiemann L, Nickel MM, Witkovsky V, Schmidt P, Gross J, Ploner M. 2015. Prefrontal Gamma Oscillations Encode Tonic Pain in Humans. Cereb Cortex 25:4407-4414.

Senkowski D, Kautz J, Hauck M, Zimmermann R, Engel AK. 2011. Emotional facial expressions

modulate pain-induced beta and gamma oscillations in sensorimotor cortex. J Neurosci 31:14542–14550.

Tiemann L, May ES, Postorino M, Schulz E, Nickel MM, Bingel U, Ploner M. 2015. Differential neurophysiological correlates of bottom-up and top-down modulations of pain. Pain 156:289–296.

Tracey I, Mantyh PW. 2007. The cerebral signature for pain perception and its modulation. Neuron 55:377–391.

Villemure C, Bushnell MC. 2009. Mood influences supraspinal pain processing separately from attention. J Neurosci 29:705–715.

Villemure C, Slotnick BM, Bushnell MC. 2003. Effects of odors on pain perception: deciphering the roles of emotion and attention. Pain 106:101–108.

Wiech K, Seymour B, Kalisch R, Stephan KE, Koltzenburg M, Driver J, Dolan RJ. 2005. Modulation of pain processing in hyperalgesia by cognitive demand. Neuroimage 27:59–69.

Zhang ZG, Hu L, Hung YS, Mouraux A, Iannetti GD. 2012. Gamma-band oscillations in the primary somatosensory cortex—a direct and obligatory correlate of subjective pain intensity. J Neurosci 32:7429–7438.

*Figure Legends*

**Figure 1.** Schematic representation of the experimental paradigm. Each trial began with a fixation cross with variable duration between 1200 and 2400 ms, followed by a picture lasting 200 ms. Then another fixation cross was presented for 1200 ms, during which painful stimuli were applied at a frequency of 3-7 Hz starting 200 ms after the prime picture. Then two consecutive scales appeared, where participants indicated the intensity and unpleasantness of the perceived painful stimuli.

**Figure 2.** Pain ratings. (A) Ratings of pain intensity and unpleasantness for each prime prime valence (negative, neutral, positive). The unpleasantness ratings showed a significant main effect of prime valence while the intensity ratings did not. (B) Across all pictures, the averaged intensity ratings were significantly positively correlated with the averaged unpleasantness ratings. VAS, visual analogue scale. * $p < 0.05$, ** $p < 0.01$. Error bars stand for standard errors.

**Figure 3.** Gamma band oscillations (GBOs). (A) Event-related spectral perturbation (ERSP) at CP2 across all pictures. The first dashed line stands for the onset of the prime stimulus and the second dashed line represents the onset of the electrical stimuli. The black rectangles indicate the time-frequency windows of the early and late GBOs. (B, C) The scalp distribution of early and late GBOs. Early GBO had a central distribution contralateral to the stimulus location and the late GBO had a centroparietal distribution. The bold black dots indicate the regions of interest used in the statistical analyses. (D, E) The ERSP value of the GBO for each prime valence. The late GBO showed a significant main effect of prime valence while the early GBO did not.

**Figure 4.** Correlations between early gamma band oscillations (GBO) and pain ratings. The mean early GBO was significantly positively correlated with (A) averaged intensity rating and (B) averaged unpleasantness rating across valence.

**Figure 5.** Correlations between late gamma band oscillations (GBO) and pain ratings. The amplitude of late GBO was positively correlated with unpleasantness rating in the negative prime condition compared with the positive prime.

**Figure 6.** The inter-trial coherence of the early gamma band oscillation (GBO) was significantly larger than that of the late GBO.